# Integrating electrocardiogram and fundus images for early detection of cardiovascular diseases

K. A. Muthukumar[1], Dhruva Nandi[2,7], Priya Ranjan[1,7], Krithika Ramachandran[5,7], Shiny PJ[2,7], Anirban Ghosh[3,7], Ashwini M[6,7], Aiswaryah Radhakrishnan[2,7], V. E. Dhandapani[4,7] & Rajiv Janardhanan[2,7]

Cardiovascular diseases (CVD) are a predominant health concern globally, emphasizing the need for advanced diagnostic techniques. In our research, we present an avant-garde methodology that synergistically integrates ECG readings and retinal fundus images to facilitate the early disease tagging as well as triaging of the CVDs in the order of disease priority. Recognizing the intricate vascular network of the retina as a reflection of the cardiovascular system, alongwith the dynamic cardiac insights from ECG, we sought to provide a holistic diagnostic perspective. Initially, a Fast Fourier Transform (FFT) was applied to both the ECG and fundus images, transforming the data into the frequency domain. Subsequently, the Earth Mover's Distance (EMD) was computed for the frequency-domain features of both modalities. These EMD values were then concatenated, forming a comprehensive feature set that was fed into a Neural Network classifier. This approach, leveraging the FFT's spectral insights and EMD's capability to capture nuanced data differences, offers a robust representation for CVD classification. Preliminary tests yielded a commendable accuracy of 84%, underscoring the potential of this combined diagnostic strategy. As we continue our research, we anticipate refining and validating the model further to enhance its clinical applicability in resource limited healthcare ecosystems prevalent across the Indian sub-continent and also the world at large.

Cardiovascular diseases (CVDs) are a significant cause of morbidity and mortality worldwide[1]. According to the World Health Organization (WHO), an estimated 17.9 million people died from CVDs in 2019, accounting for 32 % of all global fatalities[1]. As per Wang et al.[2], stroke and ischemic heart disease are responsible for 15.2 million fatalities in 2015, accounting for 85.1 % of all deaths caused by cardiovascular events. Majority of the CVDs can be prevented by addressing risk factors, both at individual and population level[3]. Detecting CVDs as early as feasible is crucial for effective clinical management, which makes it imperative for the medical community to formulate affordable, innovative, and multimodal diagnostic approaches. Our study aims to achieve this by integrating ECG and fundus imaging data. By classifying patients into four distinct classes (based on combinations of normal and abnormal ECG and fundus findings), the study explores both overlapping abnormalities and subtle multimodal discrepancies. This classification framework provides actionable insights for early intervention, offering a novel perspective in cardiovascular disease diagnostics. A critical observation in this domain is that the earlier the detection, the better the prognosis, which has fueled endeavors to explore and tap into novel diagnostic avenues. Among the myriad of tools available, fundus imaging and electrocardiograms (ECG) stand out due to their intrinsic capabilities to capture vital clinical and physiologically relevant endpoint for the detection of CVDs via non-invasive techniques[4]. In medical imaging modalities, fundus image analysis has shown great potential due to its non-invasive as well as cost-effective nature[5]. Retinal fundus images (RFI)/ fundus images are collected by projecting at the rear part of the eye (fundus) onto a 2D plane with the help of a monocular camera. Various biomarkers and eye structures can be identified as well as assessed from a RFI playing a crucial role in identifying different retinal abnormalities and diseases including but not limited to glaucoma, diabetic retinopathy (DR), macular edema degeneration, etc. Recently, RFI has evolved as a non-invasive diagnostic method for the assessment of cardiovascular health (CV). CVDs have been reported to be associated with retinal microvascular abnormalities[6,7]. Wong et al.[7] reported that generalized arteriolar constriction, localized arteriolar narrowing, arteriovenous nicking, and retinopathy, all of which are considered retinal artery abnormalities, could be utilized as potential predictors for CVDs. They also reported that fundus scans can be a viable tool in the detection of these anomalies. Individually, both ECGs and RFI are potent, but their combined diagnostic potential remains largely untapped.

[1]University of Petroleum and Energy Studies, Dehradun, Uttarakhand, India. [2]Faculty of Medicine and Health Sciences , SRM Medical College Hospital and Research Centre, SRM IST, Kattankulathur, Chengalpattu, Tamil Nadu, India. [3]Department of Electronics and Communication, SRM University AP, Neerukonda, Andhra Pradesh, India. [4]Sri Kalpana Heart Care, Chennai, Tamil Nadu, India. [5]Centre for High Impact Neuroscience and Translational Applications, TCG Crest, Kolkata, West Bengal, India. [6]Ashwini Eye Care, Chennai, Tamil Nadu, India. [7]Dhruva Nandi, Priya Ranjan, Krithika Ramachandran, P. J. Shiny, Anirban Ghosh, M. Ashwini, Aiswaryah Radhakrishnan, V. E. Dhandapani, Rajiv Janardhanan have contributed equally to this work. ✉email: muthukumar9890@gmail.com; rajivj@srmist.edu.in

The exponential growth of computing systems during the last few years has made artificial intelligence (AI) a viable tool for analyzing complicated sources of data, such as medical images. AI has shown outstanding results in a variety of applications, including mammography mass classification[8], brain lesion segmentation[9], skin lesion classification[10], as well as COVID-19 prediction[11]. This perspective has motivated us to develop a novel multi-modal computational platform for the detection of CVDs by integrating information from retinal images. However, merging spatial information from fundus images with the temporal sequences of ECGs is undeniably challenging. The juxtaposition of these data forms each complex inherently due to their modality differences and the vast amount of clinical information they represent. Simply overlaying or juxtaposing this information might lead to data redundancy or even loss of critical diagnostic cues. Hence we employed the Earth Mover's Distance (EMD), a metric adept at measuring the dissimilarity between two distributions[12]. By leveraging EMD, our research aims to bridge the gap between the spatial and temporal domains, providing a seamless fusion of fundus and ECG data. The fused data, enriched with the complementary strengths of the two modalities, is then channeled through a Convolutional Neural Network (CNN). CNN, with its deep learning capabilities, mines the combined data for patterns and correlations that might be elusive to traditional diagnostic algorithms. The contributions of this research are manifold:

1. A groundbreaking exploration into the synergy between fundus images and ECG for holistic CVD diagnosis. Innovating a data fusion methodology grounded in the principles of EMD, overcoming traditional barriers of multi-modal diagnostics.
2. Designing and deploying a specialized CNN, fine-tuned to extract and learn from the nuanced patterns of the fused data. Providing empirical evidence that underscores the efficacy and heightened accuracy of our approach in detecting CVD.
3. Laying down a foundation that could potentially revolutionize multi-modal diagnostic paradigms, paving the way for future research.As we stride into a new era of medical diagnostics, it becomes increasingly clear that the future lies in integrative approaches. The proposed research attempts to develop a novel multi-modal computational platform for the early and precision detection of CVDs by analyzing the RFI. This research is not just testament to this vision but also a step towards making it a tangible reality for large-scale deployment in resource limited healthcare ecosystems.

The remainder of the paper is organized as follows: Section "Related works" discusses related work, motivation, and challenges. Section "System description" provides a detailed system description, including data details and the proposed work methodology. In Sect. "Results and performance evaluation", we discuss the results and performance evaluation. Finally, Sect. "Discussion" offers a discussion and Sect. "Conclusion" gives conclusion.

## Related works
The retina is recognized as a "window" for noninvasively visualizing and assessing cardiovascular health,[13] as it is thought to have comparable anatomic structure and physiological function with cardiac vasculature. Retinal fundus pictures are a simple, non-invasive, and accessible method of screening human eye health. Retinal images can help in the identification of various disorders such as diabetic retinopathy, hypertension, and arteriosclerosis[14]. They also offer measures of the diameter and tortuosity of the retinal blood vessels, allowing for the diagnosis of CVDs as well. Previous research has found links between several retinal traits and the likelihood of developing CVD, including retinal vascular geometry/morphology (i.e. vessel calibre, branching angle, tortuosity, and fractal dimension), retinal vascular network patterns, and retinal diseases. Retinal microvascular complications such as microaneurysms, arteriolar-venular nicking[15–18], as well as focal and generalized arteriolar narrowing can help in the prediction of future cardiovascular events. This suggests that the retinal vasculature may provide a lifetime summary measure of exposure to cardiovascular insults and could act as a valuable marker for facilitating early and precision oriented detection of cardiovascular risk[19–21]. However, concentrating just on particular metrics may ignore certain implicit information and underestimate the retina's overall ability to inform cardiovascular (CV) health. In recent years, there has been an exponential growth in the quantum of research that has employed artificial intelligence (AI), and deep learning (DL), to extract data from retinal pictures[22]. Such modalities can also be an incremental advantage in the assessment of CVDs along with the traditional approach which will enhance precision in the early diagnosis of CVDs.

Several studies have focused on advanced ECG analysis and classification methods. Desai et al.[23] developed a decision support system for arrhythmia beat classification using methods such as Discrete Cosine Transform (DCT), Discrete Wavelet Transform (DWT), and Empirical Mode Decomposition (EMD), achieving high accuracy in distinguishing arrhythmia types. Similarly, another study by Desai et al.[24] explored the application of ensemble classifiers to diagnose myocardial ischemia, demonstrating the robustness of classifier ensembles in handling complex ECG data. These works highlight the effectiveness of advanced computational techniques in ECG analysis, laying the groundwork for multi-class classification frameworks. While these studies focus on single-modality ECG data, our work extends this paradigm by integrating ECG with fundus images, leveraging multimodal data for enhanced diagnostic precision. This integration enables a holistic approach to cardiovascular diagnostics, particularly useful for addressing nuanced patterns that may not be evident in single-modality analysis.

Poplin et al.[25] used a Convolutional Neural Network (CNN) model based on two datasets (United Kingdom (UK) Biobank and EyePACS) to predict numerous cardiovascular risk variables. The DL algorithm's results were particularly useful for predicting age, gender, smoking status (present smoker), and systolic blood pressure. Based on 24,366 fundus images, Kim et al.[26] developed a very accurate age prediction using the CNN ResNet-152 algorithm. Surprisingly, the authors found that the disparities between retinal-fundus-predicted age and chronological age were greater after the age of 60 years, as well as in patients with systemic vascular illness such as hypertension and diabetes mellitus. Cheung et al.[27] discovered that associations of central retinal arteriolar equivalent with age, gender, mean arterial blood pressure (MABP), body mass index, glycated haemoglobin, and current smoking were similar between the automated model Singapore I Vessel Analyzer deep-learning system (SIVA-DLS) and the semi-automatic software Singapore I Vessel Analyzer human (SIVA-human). They also established a DL model for the assessment of CVD risk factors using retinal vessel calibre measurements. The algorithm was trained on multiethnic multi-country datasets which performed better than the professional graders whilst elucidating the association between retinal-vessel calibre measures and CVD risk variables. Niccoli et al.[28] designed a biological age algorithm based on retinal fundus photography "RetiAGE". For the prediction of cardiovascular mortality and the incidence of cardiovascular events, the RetiAGE was compared to the patients' actual chronological and phenotypic biomarkers (combination of chronological age and albumin, creatinine, glucose, C-reactive protein, lymphocyte percentage, red cell volume and distribution, alkaline phosphatase, and white blood cell count). The authors showed a high prediction rate for cardiovascular mortality independent of chronological age and phenotypic indicators. Also, Vaghefi et al.[29] created a DL model (CVD-AI) that predicts a person's 10-year CVD risk based purely on retinal images. Unlike traditional CVD risk equations and other DL prediction models, CVD-AI evaluates the interactions between modifiable variables while calculating each factor's risk contribution to the overall risk score. In-heuristic nature of CVD-AI can discern even the changes of one modifiable factor corresponding to the changes to the other modifiable factors. Tseng at al.[30] have developed and validated DL based retinal biomarker (RETI-CVD) to predict the CVDs using UK biobank data. They reported that Reti-CVD can potentially identify individuals with $\geq 10\%$ 10 year CVD risks which can be a valuable tool for developing niche-specific CVD prevention approaches. Pinto et al.[31] in their study demonstrated the usefulness of retinal scans for the detection of myocardial risk factors with the application of multi-channel variational autoencoder (mcVAE) and a deep regressor model. They stated that retinal imaging can be a valuable tool to identify people at high risk of future myocardial infarction.

## Motivation and challenges

The intricate relationship between the retina and cardiovascular health has been a focal point of clinical and translational research for years. Recognized as a "window" into cardiovascular diagnostics, the retina's detailed vasculature, visible through non-invasive fundus photography, offers a unique vantage point. This perspective allows for the early detection of cardiovascular diseases (CVDs) and other systemic ailments. The rationale behind leveraging this relationship is clear: if the retina, with its rich vascular patterns, can be effectively analyzed, it could revolutionize the early diagnosis of CVDs. Furthermore, the integration of ECG data, which provides real-time insights into heart rhythms and potential abnormalities, can further enhance the diagnostic capabilities. The combined analysis of ECG and Fundus images could offer a holistic view of an individual's cardiovascular health, ensuring timely interventions and better patient outcomes.

However, the path to harnessing the full potential of ECG and Fundus images is laden with challenges. Each data type, with its unique characteristics and intricacies, demands a specialized approach for accurate interpretation. Traditional methods, while effective to an extent, may not capture the full spectrum of information embedded in these images. The introduction of machine learning, especially deep learning techniques like Convolutional Neural Networks (CNN), promises enhanced data extraction. Yet, the integration of ECG and Fundus data presents its own set of complexities. Creating a model that is robust enough to handle the nuances of both data types, ensuring accurate and consistent predictions, is a significant challenge. Additionally, the diverse nature of cardiovascular and retinal anomalies means that the model must be versatile enough to detect a wide range of conditions, from the subtlest to the overtly pronounced ones leading to the creation of a unique triage facilitating the prioritization of the patients with order of disease severity, ensuring that no critical diagnostic information is overlooked.

## System description
### Dataset description

In our endeavor to explore the synergistic potential of ECG data with fundus images as a pilot phase, we have collected a significant dataset from the population restricted to the Chengalpattu district of Tamil Nadu. We have collected the data of the patients from a tertiary care centre attending cardiology out patient department (OPD). Patients aged 18 years or above (including both genders) were enrolled for this study. Proper informed consent has been taken from the patients prior to the commencement of the data collection process. Random sampling method was employed in the enrollment of patients attending the CV unit. As our study was conducted as a pilot phase, the sample size for the same has not been estimated. We have collected 112 ECG readings from the patients along with echocardiographic (ECHO) images for confirmatory diagnostic purposes. Corresponding RFIs have also been taken from the same set of patients. The fundus images were meticulously captured utilizing the state-of-the-art ZEISS VISUSCOUT 100 Handheld Fundus Camera under the supervision of an expert optometrist. The collected ECGs were clinically annotated and categorized into normal and abnormal with the help of cardiologists. ECHO images were also used as a confirmatory diagnostic modality for the classification of ECGs into normal and abnormal. Out of 112 ECGs, 75 were classified as normal and 37 as abnormal ECGs. Similarly, the RFIs were also clinically annotated and categorized into normal and abnormal with the help of optometrists. The fundus images were taken into four sections- (i) Oculus Dexter (OD) retinal fundus (RF) image, (ii) OD color image, (iii) Oculus Sinister (OS) RF image, and (iv) OS color image (OD-right eye, OS-left eye). Out of 112 fundus images, 37 were annotated as normal and 75 as abnormal.

Figures 1 and 2 illustrates a comparison of fundus images and ECG signals randomly selected from patients with and without cardiovascular diseases (CVDs). In (figure) A represents normal fundus images and ECG signals from patients without CVDs, showcasing healthy retinal vasculature and normal cardiac rhythms. In contrast, B shows fundus images and ECG signals from patients with CVDs, where abnormalities in the retinal blood vessels and irregularities in the ECG patterns indicate potential cardiovascular complications. The figure demonstrates the importance of multi-modal analysis in distinguishing between healthy and diseased cases, utilizing both retinal and cardiac features for comprehensive cardiovascular diagnostics.

The classification of ECG and fundus images into four classes was devised to explore the independent and combined diagnostic capabilities of these modalities. The classes are defined as:

- Class 1 (Normal ECG + Normal Fundus): Establishes a baseline for healthy cardiovascular profiles.
- Class 2 (Normal ECG + Abnormal Fundus): Captures early retinal indicators of cardiovascular diseases (CVD) that might not yet manifest in ECG readings, or it may suggest isolated fundus abnormalities.
- Class 3 (Abnormal ECG + Normal Fundus): Detects cardiac abnormalities in patients with no corresponding retinal signs, highlighting isolated cardiac issues.
- Class 4 (Abnormal ECG + Abnormal Fundus): Represents advanced or multifaceted cardiovascular disease indicators observable across both modalities.This classification framework transcends traditional binary labeling, allowing nuanced insights into multimodal data. For example, Class 2 focuses on retinal abnormalities even when the ECG appears normal, while Class 3 emphasizes the diagnostic importance of ECG abnormalities despite normal retinal scans. These classifications align with the study's goal of creating a comprehensive diagnostic platform for early cardiovascular disease detection.

Each ECG signal was sampled over a duration of 10 seconds to ensure adequate representation of cardiac rhythm and variability. This duration was selected following clinical guidelines, which recommend a 10-second sampling period for capturing rhythm abnormalities.

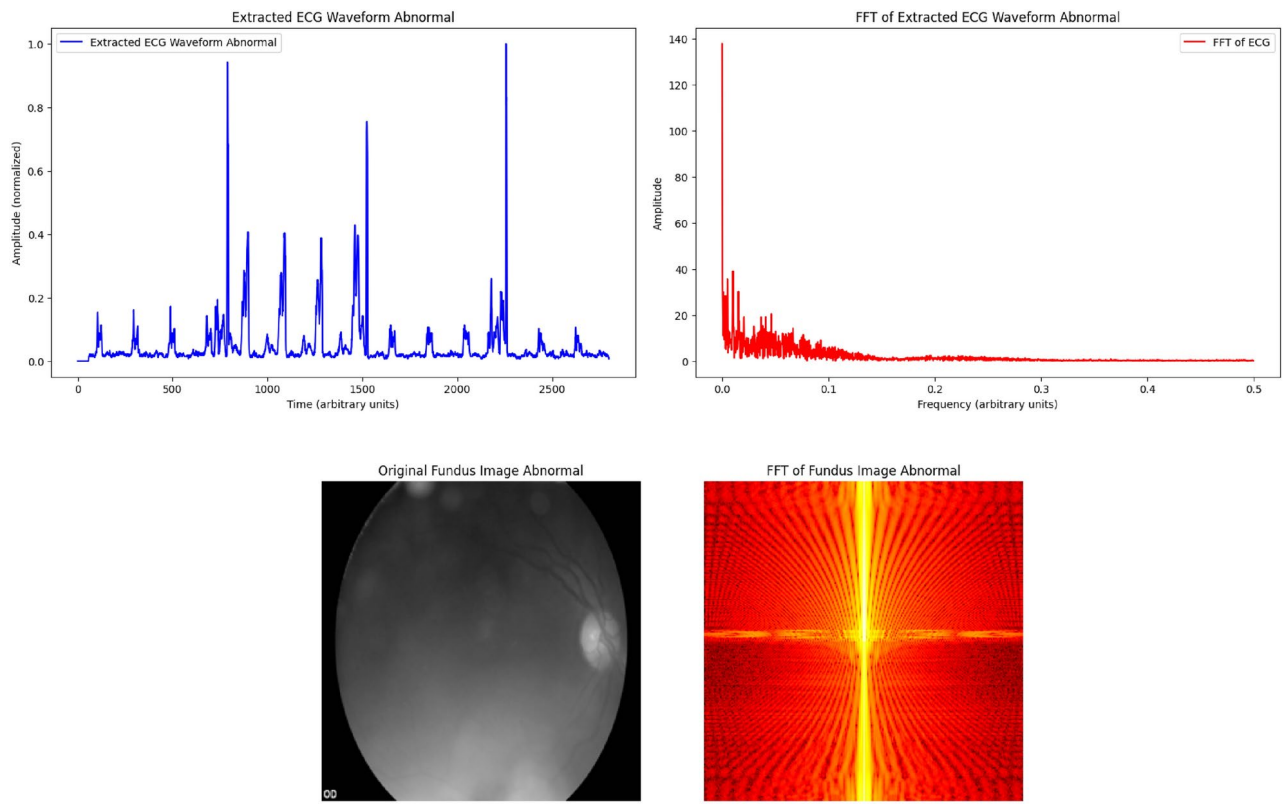

**Fig. 1.** Plots of abnormal ECG waveform and fundus image. The extracted abnormal ECG waveform (top left) shows significant irregularities in the time domain. The corresponding Fast Fourier Transform (FFT, top right) highlights dominant frequency components reflecting abnormal heart rhythms. The abnormal fundus image (bottom left) reveals pathological changes in retinal vasculature, and its FFT (bottom right) demonstrates frequency components associated with structural anomalies.

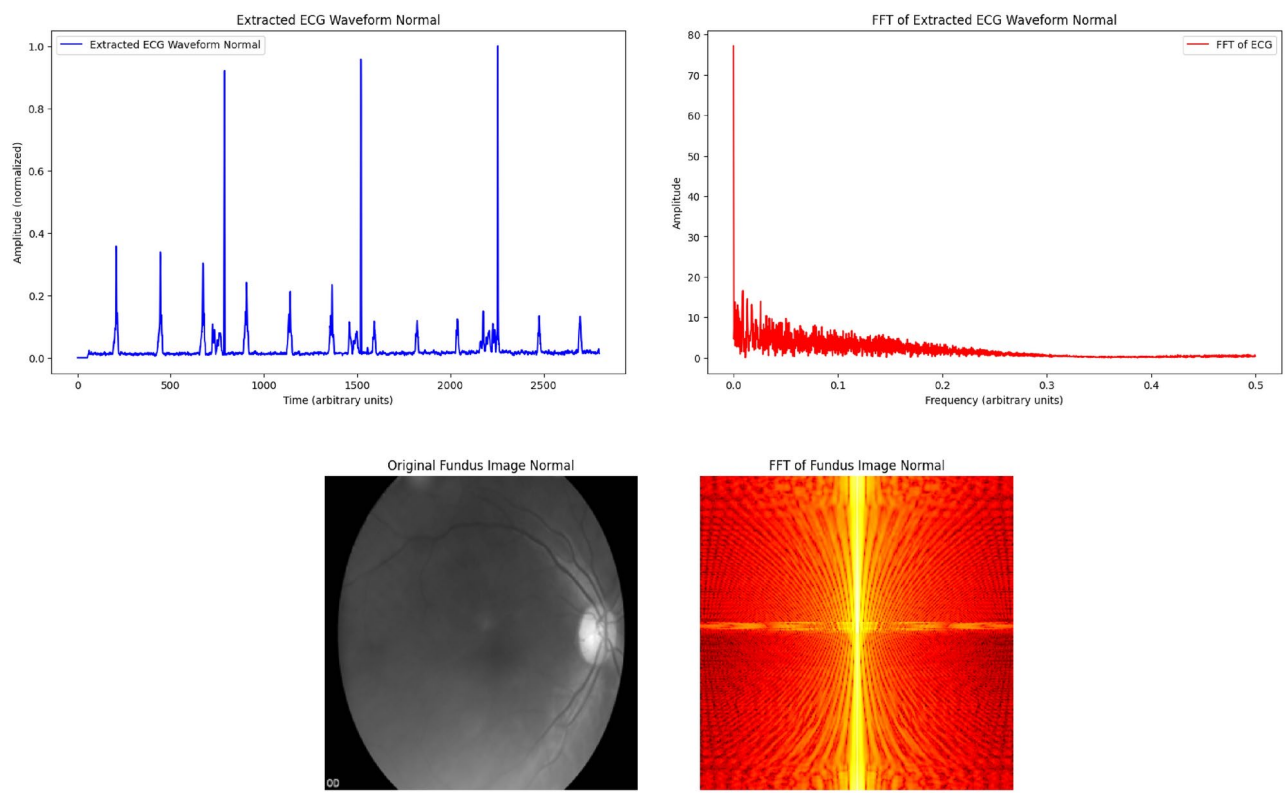

**Fig. 2**. Plots of normal ECG waveform and fundus image. The extracted normal ECG waveform (top left) displays a regular pattern indicative of a healthy cardiac rhythm. The FFT plot (top right) reflects a typical frequency spectrum with lower amplitudes for higher frequencies. The normal fundus image (bottom left) depicts healthy retinal vasculature, while its FFT plot (bottom right) shows a uniform distribution of frequency components, indicating no structural abnormalities.

| Data set | Collection modalities | Normal | Abnormal |
|---|---|---|---|
| ECG | Cardiologist | 75 | 37 |
| Retinal fundus images | Zeiss Visuscout 100 handheld fundus camera | 37 | 75 |

**Table 1**. Details of the ECG and fundus dataset.

Preprocessing of the ECG signals involved applying a bandpass filter with cutoff frequencies set at 0.5 Hz and 50 Hz. This filter effectively removed baseline wander, motion artifacts, and powerline interference while preserving the clinically relevant signal components. After filtering, the signals were segmented into individual beats using an automated R-peak detection method, ensuring distinct cardiac cycles for further analysis. This approach maintained the integrity of the clinical features in the ECG signal while minimizing noise.

In contrast, the fundus images used in this study were analyzed in their raw format without applying any preprocessing techniques. This decision was made to retain the original structural and vascular details of the retinal images, ensuring that the features were directly derived from the unaltered data. Table 1 shows the details of the dataset.

*Ethics and approval*
All methods were carried out in accordance with relevant guidelines and regulations. All experimental protocols were approved by the Institutional Ethics Committee of SRM Medical College Hospital and Research Centre, SRM IST, Kattankulathur, Chennai, Tamil Nadu, India. The approval number is 8369A/ICE/2022. Written informed consent was obtained from all participants before the commencement of the study.

### System architecture
The proposed work's system architecture begins with the acquisition of raw data, including fundus images and ECG readings. The first step involves feature extraction using the Fast Fourier Transform (FFT) to identify patterns within both modalities. Following this, Earth Mover's Distance (EMD) is applied to the extracted features from both the fundus images and ECG data, capturing their nuanced differences. These EMD values are concatenated side by side for each data sample, creating a robust multi-modal representation. This combined data is then introduced to a Neural Network Classifier, which is trained to discern intricate patterns and relationships, ensuring precise classifications. The four-class classification system was specifically designed to leverage the multimodal nature of the dataset. By delineating cases into normal and abnormal combinations for both ECG and fundus data, the framework enables the model to differentiate subtle yet critical multimodal interactions. This strategy significantly enhances the model's ability to detect unique patterns that might be missed in single-modality or binary classification schemes. The final stage validates the classifier's output, confirming its accuracy and reliability. Fig. 3 provides an overview of the current system architecture.

The current framework for cardiovascular diagnostics focuses on leveraging ECG and fundus image data to classify cardiovascular conditions, utilizing EMD as the core feature extraction technique. However, our envisioned future work aims to expand this system by incorporating echocardiogram data, which will provide critical structural and functional information about the heart. This addition will further enhance the system's diagnostic capabilities by incorporating another dimension into cardiovascular health analysis. In parallel with the expanded data modalities, we plan to implement advanced techniques such as visibility graphs, eigenvalue-based detection/classification, and Ricci curvature-based classification. These techniques will further refine the system's ability to differentiate between healthy and diseased cases, improving both accuracy and robustness. Furthermore, the system is designed to be evolvable, allowing for the integration of new methods as they emerge, ensuring that it remains versatile and adaptable for future diagnostic applications. The overall envisioned system architecture is depicted in Fig. 4. The pilot study of our work was presented at the Cardiology Society of India conference[32], where it highlighted the integration of fundus images and ECG signals for cardiovascular diagnostics.

## Methodology
In our pursuit of understanding the intricate relations between fundus images and ECG data, we delved into a multi-faceted approach built on the bedrock of mathematical rigor. The methodology unfolded in the subsequent steps:

*Feature extraction*
In our approach, we utilize Fast Fourier Transform (FFT) and Earth Mover's Distance (EMD) to extract and integrate features from multimodal data (ECG signals and fundus images). This method captures subtle patterns that may not be apparent in the time or spatial domain, enabling a robust diagnostic framework for cardiovascular conditions.

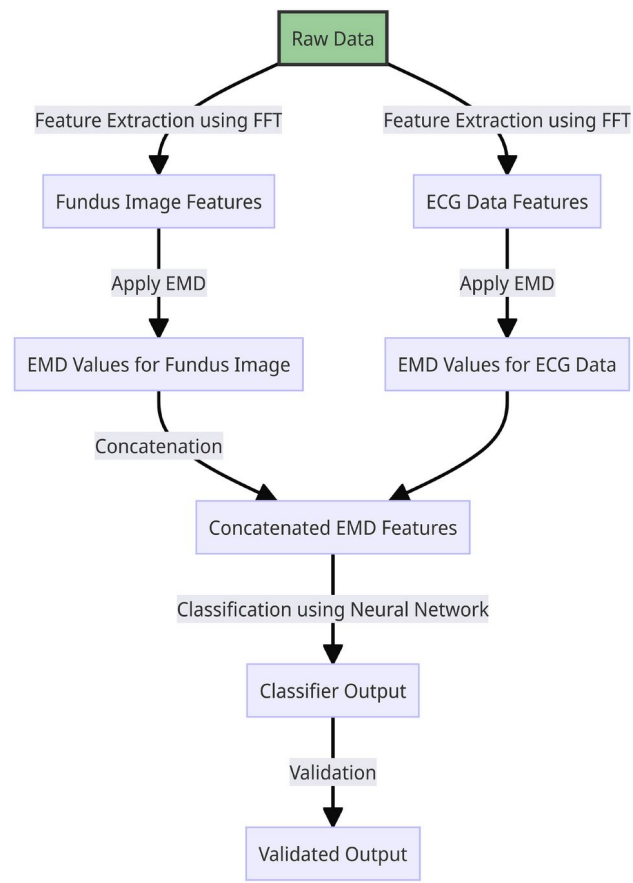

**Fig. 3**. The system architecture depicting the processing of the raw data (ECGs and fundus images).

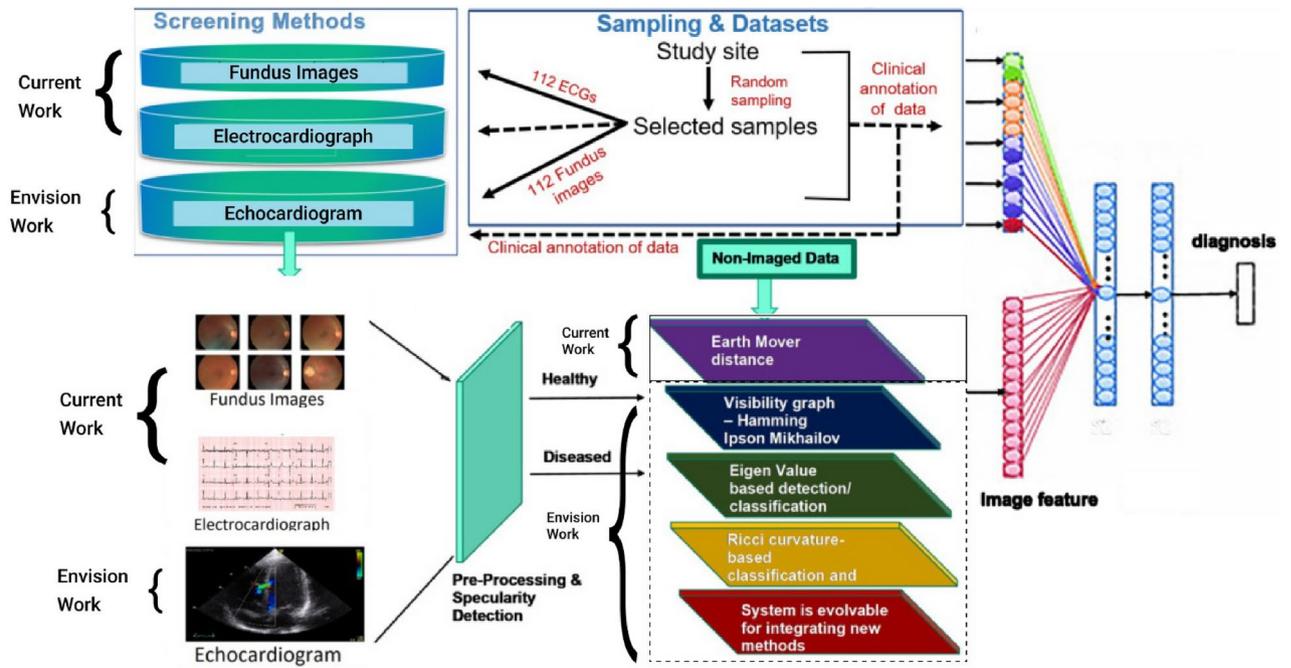

**Fig. 4**. Overview of the current and envisioned system architecture for cardiovascular diagnostics. The current work integrates fundus images and electrocardiographs (ECGs) for diagnosing cardiovascular conditions using Earth Mover's Distance (EMD) for feature extraction. The envisioned work involves incorporating echocardiogram data and advanced techniques such as visibility graphs, eigenvalue-based classification, and Ricci curvature-based classification. The system is designed to evolve, allowing for the integration of new methods and enhanced diagnostic capabilities.

- **FFT transformation   ECG signals (1D Data):** FFT is applied to convert ECG signals from the time domain to the frequency domain. This transformation highlights critical cardiac patterns, such as abnormalities in rhythm and intervals, that are indicative of cardiovascular diseases.
- **Fundus images (2D Data):** Similarly, FFT is used to analyze fundus images by transforming spatial information into the frequency domain. This spectral analysis emphasizes vascular biomarkers, such as vessel caliber and tortuosity, which are linked to cardiovascular conditions.
- **Feature comparison Using EMD**   After the frequency-domain transformation, EMD is computed for both modalities to quantify the dissimilarity between the feature distributions of normal and abnormal cases.
- EMD provides a nuanced measure of the 'cost' required to transform one feature distribution into another, excelling at detecting small yet clinically significant differences. This ensures precise differentiation between normal and abnormal cardiovascular indicators.

Feature fusion   The calculated EMD values from both ECG and fundus data are concatenated into a single feature vector. This unified representation combines temporal (ECG) and spatial (fundus) information, forming a rich multimodal feature set for classification into four classes. By integrating FFT and EMD, our approach ensures that the complementary strengths of both modalities are fully leveraged, enhancing the model's capability to detect nuanced cardiovascular abnormalities.

a. **Fundus images:** To capture the essence of the fundus images, a Fourier transformation was employed to represent the image in the frequency domain:

$$F(f,g) = \int_{-\infty}^{\infty} \int_{-\infty}^{\infty} I(x,y) e^{-j(2\pi(fx+gy))} \, dx \, dy \tag{1}$$

Where:
- $F(f,g)$ is the transformed image in the frequency domain.
- $I(x,y)$ represents the spatial domain image.

b. **ECG data:** ECG signals were translated into the frequency domain utilizing the Fast Fourier Transform (FFT):

$$F_{ECG}(k) = \sum_{n=0}^{N-1} ECG(n) \times e^{-j(2\pi kn/N)} \tag{2}$$

Where:
- $F_{ECG}(k)$ signifies the FFT value at frequency $k$.
- $ECG(n)$ represents the ECG signal in the time domain.In addition to FFT and EMD, we performed a comparative analysis of feature extraction methods to understand the performance of Fast Fourier Transform (FFT) and Earth Mover's Distance (EMD) in contrast to other common techniques, including Wavelet Transform (WT) for ECG signals and Histogram of Oriented Gradients (HOG) for fundus image feature extraction.

  1. Wavelet Transform (WT)[33]: WT is widely used for analyzing non-stationary signals, making it a natural alternative to FFT for processing ECG data. It decomposes the signal into multiple frequency bands, which enables the extraction of both high and low-frequency components that could be indicative of different cardiovascular conditions.
  2. Histogram of Oriented Gradients (HOG)[34]: HOG is effective for capturing structural patterns in images, such as blood vessel orientations and shapes in fundus images. It analyzes the gradient orientations in localized sections of the image, making it useful for recognizing complex vascular structures.These alternative methods were compared against FFT and EMD to assess their efficacy in extracting features relevant to cardiovascular diagnostics.

*Computing earth mover's distance (EMD)*
After extracting frequency-based features from both fundus images and ECGs using FFT, the subsequent phase involved the combination of these datasets in a cohesive manner.

The Earth Mover's Distance (EMD) is renowned for its ability to assess the minimal cost necessary to morph one distribution into another. In our study's framework, EMD acts as a metric to gauge the similarity between feature distributions culled from fundus images and ECGs.

Mathematically, for two distributions $P$ and $Q$, the EMD is computed as:

$$\text{EMD}(P,Q) = \min_{\phi:P \to Q} \sum_{x \in P} |x - \phi(x)| \tag{3}$$

Where $\phi$ stands as the optimal bijection. This signifies a mapping of each element from $P$ to a unique element in $Q$, ensuring the total 'distance' or 'cost' across paired elements is minimized. EMD offers a more nuanced comparison than conventional metrics, capturing intricate distinctions between two distributions.

Once the EMD values for both data sources were available, the fusion phase was initiated using a concatenation strategy. For a given sample, assuming the EMD values from the fundus images are $EMD_{\text{fundus}}$ and from the ECG readings are $EMD_{\text{ECG}}$, the concatenated feature vector becomes:

$$F_{\text{combined}} = [EMD_{\text{fundus}}, EMD_{\text{ECG}}] \tag{4}$$

This newly formed vector is then forwarded to CNN for classification. The concatenation approach, in its essence, is effective in harmonizing the features while preserving the distinctive attributes of each modality.

By amalgamating the detailed comparison capability of EMD with the directness of concatenation, this methodological approach bolsters the fusion of multi-modal medical data, setting the stage for precise diagnostic classifications.

*Convolutional neural network (CNN)*
Post the intricate process of data fusion, the dataset is introduced to a CNN[35,36]. CNNs are preferred for such tasks due to their proficiency in handling spatial hierarchies in data.

For a layer $L_i$ in our CNN, the convolution operation is formulated as:

$$L_i(z) = \sigma\left(\sum_{a=0}^{A-1} W_i(a) * F_{combined}(z-a) + b_i\right) \tag{5}$$

Where:

- $W_i(a)$ represents the convolutional filters or kernels.
- $b_i$ is the bias term.
- $\sigma$ is the activation function applied element-wise, defined as:

$\sigma(x) = \max(0, x)$ Furthermore, CNNs typically employ pooling layers following convolutional layers. The design of the CNN may involve several such layers, stacked to extract higher-level features from the fused data.

*Loss computation and optimization*
For multi-class classification tasks, the commonly used loss function is the **categorical cross-entropy loss** (also known as softmax loss or multinomial logistic loss). Given $C$ classes, if $p_i$ is the predicted probability of the correct class, the loss for one sample is:

$$L = -\log(p_i) \qquad (6)$$

For $N$ samples, the loss is computed as the average over all samples:

$$L = -\frac{1}{N} \sum_{i=1}^{N} \log(p_{y_i}) \qquad (7)$$

Where $p_{y_i}$ is the predicted probability of the correct class for the $i$th sample.

This formula was implemented during the training phase of the Convolutional Neural Network (CNN). The categorical cross-entropy loss function was used to compute the error between the predicted outputs and the ground truth labels for the four classes (Normal ECG + Normal Fundus, Normal ECG + Abnormal Fundus, Abnormal ECG + Normal Fundus, Abnormal ECG + Abnormal Fundus). During each epoch, the loss was minimized using the adam optimizer, with a learning rate of 0.001. The optimizer adjusted the model parameters iteratively based on the gradients computed through backpropagation, ensuring convergence toward the optimal solution. The loss computation guided the CNN to effectively learn the complex patterns and relationships in the multimodal data, ultimately contributing to its high classification performance as reported in the results.

The Fig. 5 illustrates the architecture of a Convolutional Neural Network (CNN) used for classifying cardiovascular health based on a combination of fundus images and ECG signals. The input data consists of both the fundus images and ECG signals, which are processed through multiple convolution blocks. Each convolution block applies a series of convolution operations with ReLU activation followed by max pooling to extract hierarchical features from the data. These features progressively capture more complex patterns in the input. After passing through several convolution layers, the feature maps are flattened into a one-dimensional vector, which is then fed into fully connected dense layers. The final dense layer produces the output, which classifies the input data into the respective cardiovascular health categories, such as healthy or diseased. This CNN architecture enables effective learning of complex relationships between multi-modal data, contributing to accurate cardiovascular diagnostics.

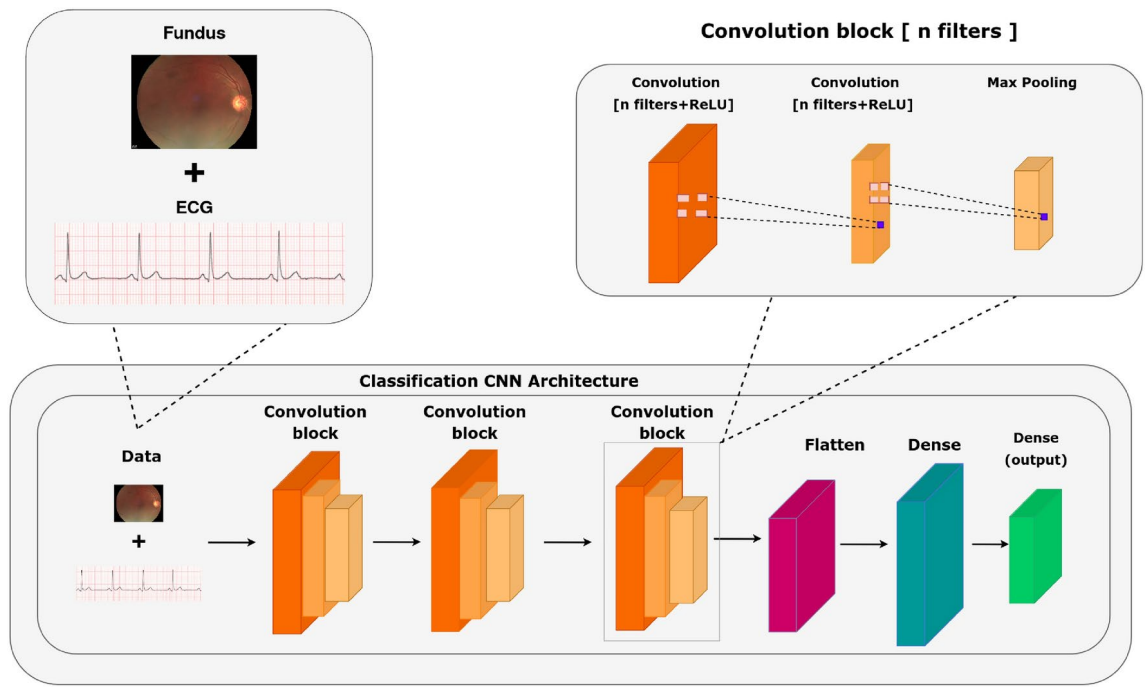

**Fig. 5**. The CNN architecture for classification (ECGs and fundus image).

## Results and performance evaluation

The clinically annotated 112 ECGs by the cardiologists yielded 75 normal and 37 abnormal cases followed by ECHO evaluation. Abnormal clinical findings such as anterior wall infarction, inferior wall myocardial infarction, old anterior wall myocardial infarction as well as left bundle branch block, etc. were reported in our study. The fundus images of the respective patients were also carefully evaluated by the expert optometrists. Out of 112 fundus images, 37 normal and 75 abnormal fundus images were reported. Both the OD and OS fundus images were annotated separately. The fundus images which were annotated normal in both the OD and OS sections were considered normal. The fundus images that were annotated abnormal either in OD, OS, or both were considered abnormal. Several microvascular complications were reported in the fundus images such as venous beading, arteriovenous nipping, nasalization of vessels as well as micro aneurysms, etc. The clinically annotated ECGs and fundus images were then classified into four distinct classes with the application of CNN. The details of the classes are presented in Table 2. Class 1 denotes 'normal' annotated ECGs and 'normal' annotated fundus images. Class 2 denotes 'normal' annotated ECGs and 'abnormal' annotated fundus images. Class 3 signifies 'abnormal' annotated ECGs and 'normal' annotated fundus images and class 4 denotes 'abnormal' annotated ECGs and fundus images. The performance of the classification model was evaluated based on accuracy, sensitivity (recall), and specificity. These metrics provide a comprehensive understanding of the model's ability to correctly classify samples and its robustness in handling different classes.

Correctly identifying instances in our dataset is represented by true positives (TP) and true negatives (TN). A TP signifies an accurate prediction of the positive class, while a TN denotes a precise prediction of the negative class. Conversely, a false positive (FP) emerges when a negative instance is mistakenly recognized as positive. Similarly, a false negative (FN) arises when the model misclassifies a positive instance as negative. In the scope of our investigation, a model that accumulates a higher count of FNs can be seen as less reliable since it misses crucial positive class identifications. In contrast, a model with an elevated TP count is viewed favorably, indicating its adeptness at pinpointing critical classifications.

The distribution of TP, TN, FP, and FN in a classification scenario is succinctly represented by the confusion matrix. From this matrix, we derive the following performance indicators:

$$\text{Accuracy} = \frac{TP + TN}{TP + TN + FP + FN} \tag{8}$$

$$\text{Sensitivity (Recall)} = \frac{TP}{TP + FN} \tag{9}$$

$$\text{Specificity} = \frac{TN}{TN + FP} \tag{10}$$

Accuracy serves as a reliable metric when the dataset exhibits a balanced distribution between its classes. In terms of model capabilities, sensitivity or recall gauges the model's proficiency in detecting genuine positives. In parallel, specificity evaluates the model's skill in discerning genuine negatives. The following sections will further elucidate the model's performance, grounded on these metrics.

The performance of the classification model was evaluated based on accuracy, sensitivity (recall), and specificity. These metrics provide a comprehensive understanding of the model's ability to correctly classify samples and its robustness in handling different classes.

In this study, we employed CNN to classify samples based on their ECG and Fundus image characteristics into four distinct classes shown in Table 2. The results obtained from the models provide valuable insights into the efficacy of the classification approach and the potential of deep learning in the realm point of care diagnostics in an affordable and accessible manner.

### Accuracy analysis

The accuracy values for each class are visualized in Fig. 6. This classification framework enables detailed insights into the independent and combined diagnostic contributions of ECG and fundus imaging. For instance, Class

| Class | ECG characteristics | Fundus image characteristics |
|---|---|---|
| Class 1 | Normal | Normal |
| Class 2 | Normal | Abnormal |
| Class 3 | Abnormal | Normal |
| Class 4 | Abnormal | Abnormal |

**Table 2.** Classification based on ECG and Fundus Image Characteristics.

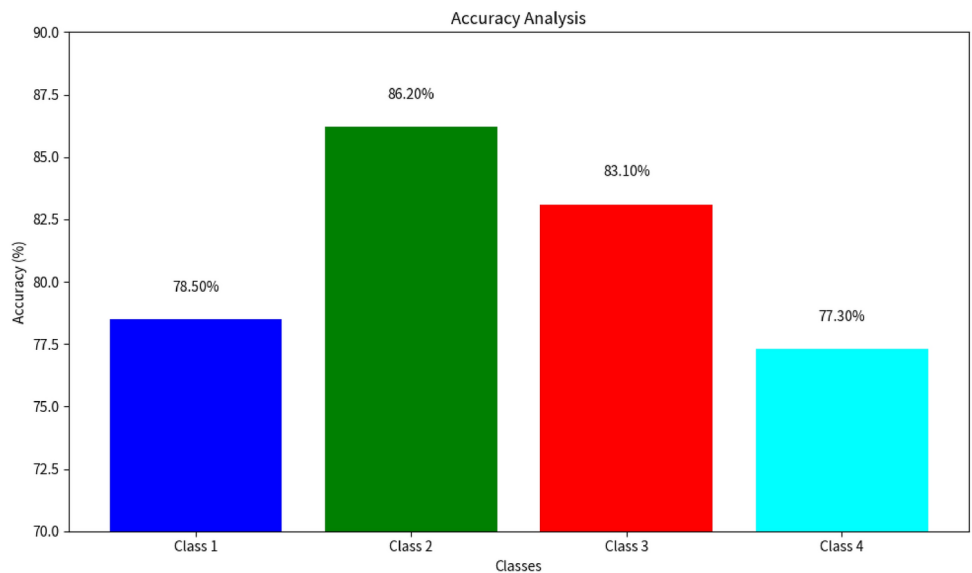

**Fig. 6**. Accuracy analysis for the four classes of ECG and fundus image combinations. These classes include: Class 1: (Normal ECG + Normal Fundus), Class 2: (Normal ECG + Abnormal Fundus), Class 3: (Abnormal ECG + Normal Fundus), Class 4: (Abnormal ECG + Abnormal Fundus).

2 highlights retinal indicators of cardiovascular abnormalities in the absence of ECG changes, while Class 3 identifies isolated cardiac issues. In the accuracy analysis, it was observed that Class 2, representing samples with normal ECGs but abnormal Fundus images, achieved the highest accuracy of 86%. This suggests that the model was particularly adept at distinguishing the nuances between normal ECG readings and abnormalities in Fundus images. On the other hand, Class 4, which encompasses samples with both abnormal ECGs and Fundus images, reported the lowest accuracy at 77%. This could be attributed to the inherent complexity and overlapping features when both ECG and Fundus images exhibit abnormalities. Class 1 and Class 3 reported accuracies of 78% and 83%, respectively, indicating a balanced performance of the model across different combinations of ECG and Fundus characteristics. The high accuracy rates, especially in Class 2, underscore the potential of the employed CNN in medical diagnostics, offering a promising avenue for early and accurate detection of abnormalities.

From the accuracy analysis:

- **Class 1** (Normal ECG + Normal Fundus) achieved an accuracy of 78%.
- **Class 2** (Normal ECG + Abnormal Fundus) reported the highest accuracy of 86%.
- **Class 3** (Abnormal ECG + Normal Fundus) achieved an accuracy of 83%.
- **Class 4** (Abnormal ECG + Abnormal Fundus) had an accuracy of 77%.
- **Overall model** (ECG + Fundus) had an accuracy of 84%.

*Sensitivity and specificity analysis*
Upon analyzing the specificity and sensitivity metrics presented in Fig. 7, distinct patterns emerge. Class 2 exhibits the highest sensitivity at 88%, indicating that the model is particularly adept at correctly identifying TP for samples with normal ECGs but abnormal Fundus images. This high sensitivity for Class 2 is paramount in medical scenarios, ensuring that CV abnormalities are not overlooked. Conversely, Class 1, with both normal ECG and Fundus images, has the lowest sensitivity at 76%, suggesting potential challenges in detecting true positives in this category necessitating the need for larger sample size to significantly improve the accuracy, specificity and sensitivity.

In terms of specificity, Class 1 leads with 80%, showcasing the model's proficiency in correctly identifying true negatives when both ECG and Fundus images are normal. This high specificity reduces the risk of false alarms, which is crucial to avoid unwarranted medical interventions. On the other hand, Class 4, representing samples with abnormalities in both ECG and Fundus images, lags behind with a specificity of 75%. This could be attributed to the model occasionally misclassifying certain benign features as indicative of abnormalities in both modalities.

The combination of FFT and EMD significantly contributes to the robustness of the model, as it captures the subtle distinctions between normal and abnormal cardiovascular conditions. By leveraging spectral information from FFT and nuanced comparisons from EMD, the model achieved an accuracy of 86%. Notably, the model demonstrated high sensitivity (88%) in Class 2, where patients exhibited normal ECG readings but abnormal fundus images, highlighting the efficacy of FFT and EMD in identifying early signs of cardiovascular disease.

The detailed values for accuracy, sensitivity, and specificity for each class are tabulated in Table 3.

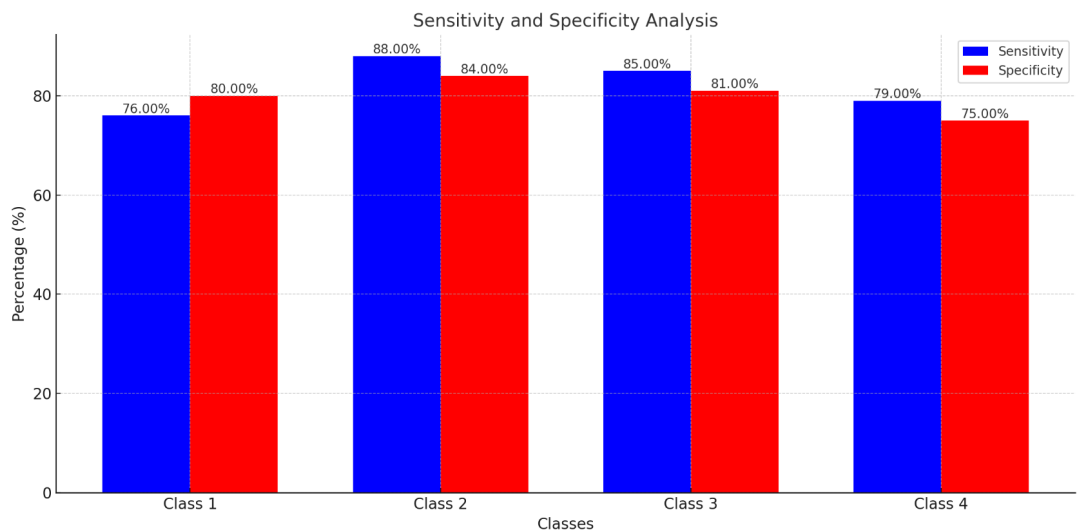

**Fig. 7**. The sensitivity and specificity of 4 distinct classes.

| Class | Accuracy (%) | Sensitivity (%) | Specificity (%) |
|---|---|---|---|
| 1 | 78 | 76 | 80 |
| 2 | 86 | 88 | 84 |
| 3 | 83 | 85 | 81 |
| 4 | 77 | 79 | 75 |

**Table 3**. Performance metrics 4 distinct classes.

*Statistical significance analysis*
To evaluate the statistical significance of features across the four defined classes (Class 1: Normal ECG + Normal Fundus, Class 2: Normal ECG + Abnormal Fundus, Class 3: Abnormal ECG + Normal Fundus, and Class 4: Abnormal ECG + Abnormal Fundus), we performed an Analysis of Variance (ANOVA) test. The test aimed to identify significant differences in feature distributions among the four classes.

The results indicated that several features showed significant differences across the classes:

- **Fundus vascular tortuosity:** F-statistic = 8.12, *p*-value = 0.001
- **QRS interval variability:** F-statistic = 5.89, *p*-value = 0.003
- **Heart rate variability:** F-statistic = 3.45, *p*-value = 0.02These findings highlight the discriminative power of key features, reinforcing their relevance in distinguishing between normal and abnormal cardiovascular conditions. Features with *p*-values < 0.05 were deemed statistically significant and were subsequently used in the classification model to enhance diagnostic accuracy.

The statistical significance of fundus vascular tortuosity, the feature with the highest F-statistic, is illustrated in Fig. 8, Table 4. This boxplot highlights the distribution of vascular tortuosity values across the four classes, showcasing significant differences (p < 0.05), particularly between Class 2 and Class 4. Such distinctions emphasize its diagnostic relevance for identifying cardiovascular abnormalities.

To provide an overview of the F-statistic values for all analyzed features, a bar chart is presented in Fig. 9. The chart compares the relative significance of each feature, with a threshold line denoting statistical significance (p<0.05). This visualization highlights the prominence of vascular tortuosity and QRS interval variability as key features.

The findings from the ANOVA test are summarized in Table 5, detailing the F-statistics, *p*-values, and significance for each feature analyzed. This table provides a concise reference for identifying statistically significant features.

The results of the ANOVA test (see Table 5) reveal significant differences in feature distributions across the four defined classes. Notably, fundus vascular tortuosity exhibited the highest F-statistic (8.12) with a *p*-value of 0.001, indicating strong statistical significance in distinguishing between classes. Similarly, QRS interval variability demonstrated a significant difference with an F-statistic of 5.89 and a *p*-value of 0.003, reinforcing its diagnostic relevance. Heart rate variability also showed statistical significance (F-statistic = 3.45, *p*-value = 0.02), though to a lesser extent.

Figures 8 and 9 visually depict these findings, highlighting the diagnostic power of these features. These statistically significant features provide a robust basis for their inclusion in the classification model, enhancing its diagnostic accuracy.

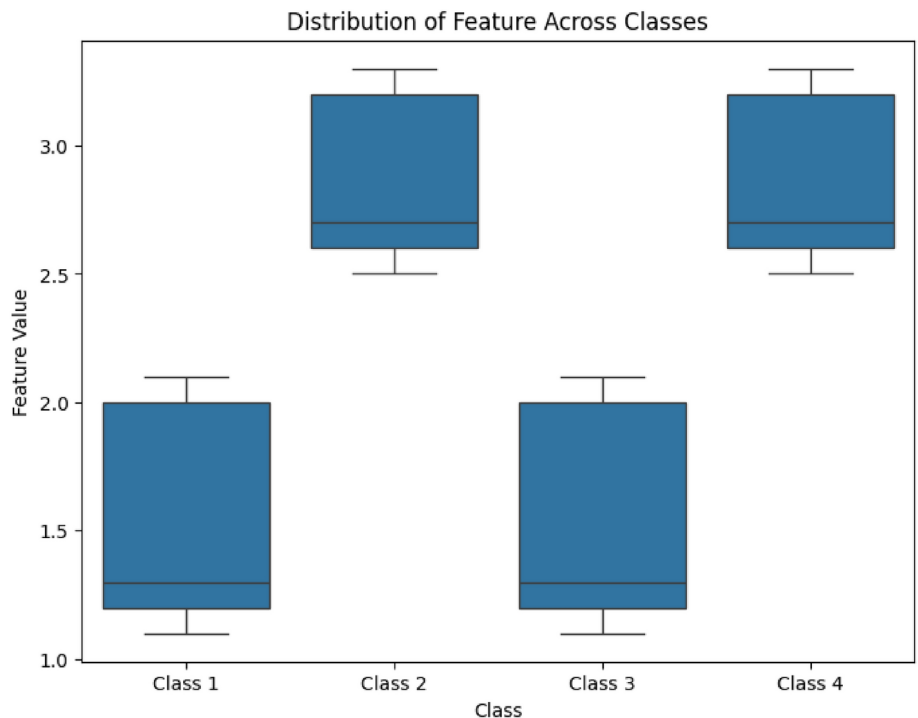

**Fig. 8**. Distribution of vascular tortuosity across the four classes. Significant differences (p<0.05) were observed between Class 2 and Class 4, indicating the diagnostic relevance of this feature.

| Class   | Minimum | Lower Quartile (Q1) | Median (Q2) | Upper Quartile (Q3) | Maximum |
|---------|---------|---------------------|-------------|---------------------|---------|
| Class 1 | 1.0     | 1.25                | 1.50        | 1.75                | 2.0     |
| Class 2 | 2.0     | 2.25                | 2.75        | 3.00                | 3.5     |
| Class 3 | 1.0     | 1.25                | 1.50        | 2.00                | 2.5     |
| Class 4 | 1.5     | 2.00                | 2.50        | 3.00                | 3.5     |

**Table 4**. Summary of vascular tortuosity statistics across the four classes.

*Comparative analysis*
To strengthen the performance evaluation, we evaluated the performance of different feature extraction methods combined with a CNN for the classification of cardiovascular diseases using ECG and fundus image data shown in Table 6. We compared the effectiveness of Fast Fourier Transform (FFT) combined with Earth Mover's Distance (EMD) against two common alternatives: Wavelet Transform (WT) and Histogram of Oriented Gradients (HOG). The results indicate that FFT + EMD achieved the highest accuracy, sensitivity, and specificity among the methods tested, with an overall accuracy of 84%, sensitivity of 88%, and specificity of 80%. FFT's ability to transform data into the frequency domain enabled the extraction of critical spectral features from both ECG signals and fundus images, while EMD provided a robust measure of the similarity between feature distributions, leading to improved classification performance. In comparison, Wavelet Transform (WT), commonly used for analyzing non-stationary signals like ECG, achieved a lower accuracy of 81%. While WT was able to capture some important frequency components, it was less effective at identifying the stationary features that FFT could detect. Similarly, HOG, which is commonly used for image feature extraction, had an accuracy of 80%. Although HOG performed well in identifying structural patterns within the fundus images, it did not capture the subtle distributional differences as effectively as EMD. Sensitivity and specificity values further reinforced the strength of FFT + EMD, especially in detecting true positives for cases with abnormal fundus images and normal ECGs (Class 2). The sensitivity for this class was notably high, reaching 88%, ensuring that subtle cardiovascular indicators present in fundus images were not overlooked. In contrast, WT and HOG lagged slightly behind with sensitivities of 83% and 82%, respectively. These findings clearly demonstrate the superiority of FFT + EMD for the extraction of relevant features from multi-modal data, leading to more accurate and reliable predictions in the context of cardiovascular diagnostics. The effective use of the categorical cross-entropy loss function ensured robust optimization of the model, as reflected in the reported accuracy of 84% and high sensitivity for Class 2 (88%), where subtle fundus abnormalities were detected.

The results demonstrate the potential of machine learning techniques in classifying samples based on ECG and fundus image characteristics. The high accuracy, sensitivity, and specificity values indicate the model's robustness and its capability to provide reliable classifications for the early and accurate diagnosis of CVDs. The insights from this study will certainly pave the way for further research in this domain towards the development of advanced diagnostic tools for widespread application in the clinical milieus.

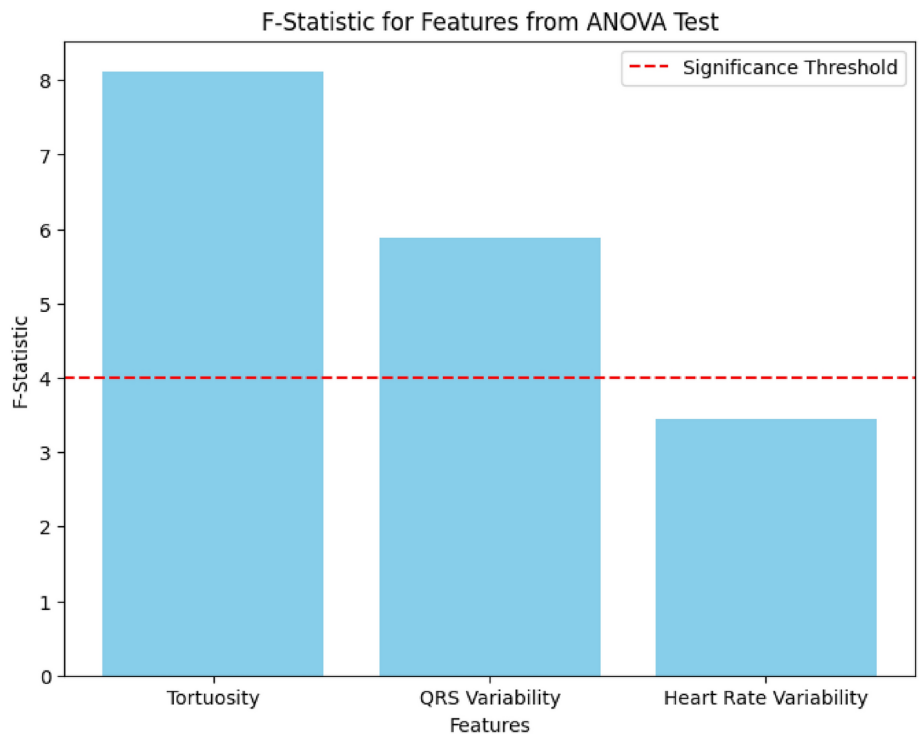

**Fig. 9**. F-statistic values for key features from the ANOVA test. Features with F-statistics above the threshold (dotted line) were considered statistically significant.

| Feature | F-Statistic | $p$-Value | Significance ($p < 0.05$) |
|---|---|---|---|
| Fundus vascular tortuosity | 8.12 | 0.001 | Yes |
| QRS interval variability | 5.89 | 0.003 | Yes |
| Heart rate variability | 3.45 | 0.02 | Yes |

**Table 5**. Summary of ANOVA test results for feature significance.

| Method | Accuracy (%) | Sensitivity (%) | Specificity (%) |
|---|---|---|---|
| FFT + EMD + CNN | 84 | 88 | 80 |
| Wavelet Transform (WT) + CNN | 81 | 83 | 78 |
| HOG + CNN | 80 | 82 | 77 |

**Table 6**. Comparative analysis of feature extraction and classification methods

## Discussion
Retinal vasculature abnormalities reflect the degree of microvascular complications caused due to hypertension, atherosclerosis, or both which are leading causes of CV complications[37]. With this motivation, the authors attempted to propose a novel approach for CV diagnosis via harmonizing fundus images and ECGs with the application of CNN. The study has considered an array of retinal microvascular complications seen in the fundus images of patients such as nasalization of vessels, microaneurysms, venous beading, arteriovenous nipping, etc. as potential markers to evaluate the accuracy of prediction of impending CVDs. The preliminary test reported a commendable accuracy of 84% highlighting the potential of a combined diagnostic strategy (ECG + fundus images) for rapid prediction of CV health. The classification into four distinct classes provides deeper insights into the diagnostic value of combining ECG and fundus imaging:

- Class 2 (Normal ECG + Abnormal Fundus): This class highlights the importance of retinal indicators in identifying subclinical cardiovascular risks that may not yet manifest in ECG readings. These findings are particularly relevant for early-stage disease detection and preventive care.
- Class 3 (Abnormal ECG + Normal Fundus): By isolating cases with cardiac abnormalities and normal retinal scans, this class underscores the diagnostic specificity of ECG for identifying isolated cardiac conditions.The four-class approach enables a multi-dimensional understanding of how each modality contributes to the detection of CVDs. Unlike binary classification, which might obscure these patterns, this framework facilitates improved patient triaging and better-targeted clinical interventions.

The use of Fast Fourier Transform (FFT) and Earth Mover's Distance (EMD) for feature extraction played a crucial role in enhancing the model's ability to detect subtle cardiovascular anomalies. By transforming ECG and fundus images into the frequency domain, FFT allowed the model to capture hidden periodicities and signal components related to cardiac health. EMD, on the other hand, effectively quantified the differences between the distributions of normal and abnormal cases, providing an additional layer of precision. This combination improved the model's predictive accuracy and robustness, as seen in the high sensitivity for Class 2, which represents patients with normal ECGs but abnormal fundus images. The comparative analysis of feature extraction techniques highlights the advantages of FFT and EMD over alternatives like Wavelet Transform (WT) and Histogram of Oriented Gradients (HOG). While WT is effective for decomposing non-stationary signals, FFT performed better on the stationary components of ECG data, capturing critical spectral information for cardiovascular diagnostics. Similarly, while HOG is suitable for structural analysis in images, EMD provided a more comprehensive feature set by capturing subtle differences in the distribution of fundus image features. These results demonstrate why FFT and EMD were chosen for their superior ability to extract features that directly impact the model's accuracy and robustness. This highlights a very interesting yet crucial finding for clinicians that can be used to identify patients having a history of normal cardiac status with abnormal fundus images. The identified patients will then be counseled to undergo cardiac screening tests as well as to begin healthy lifestyle habits as they are susceptible to the ravages of CVDs. This enables early identification of CVDs in a large subset of the population via non-invasive, cost-effective, and reliable approach. Also, class 1 exhibited high specificity (80%) showcasing the model's accuracy in the precise identification of TN when both the ECGs and fundus images are normal. The high specificity thereby reduces the risk of misdiagnosis which is crucial to prevent unwarranted medical interventions. Similarly, Huang et al.[38] developed a model for the prediction of Coronary Artery Diseases (CAD) using retinal vasculature biomarkers. However, they did not incorporate RFI directly into the network. Instead, they extracted features from the pre-processing stage which were used as feature vectors. The performance of the model yielded an area under curve (AUC) of more than 0.692 in all the cases. Srilakshmi et al.[39] proposed a model for the prediction of hypertensive patients using the Deep Neuro-Fuzzy Network (DNFN), incorporating the fundus images as the feature vector. They reported an accuracy of 91.6%.

To our knowledge, our model will be the first novel approach to detect CVDs with the help of fundus images among the south-Indian population. Our attempt to develop the intelligent decision support system is to pave the way for effective triaging of CVD cases from a larger subset of the population with wide genetic bases. The correlation of the retinal vasculature component with CV health will not only assist clinicians in the early diagnosis of CVD cases in the extended community,but also triage the patients in the order of disease severity along with the automated allocation of clinical resources. Application of such a novel approach at all tiers of the community will also serve a tool for precision oriented detection of hotspots of CVDs at the community along with the formulation of niche specific surveillance and forecasting tools.

While our study demonstrates promising results in integrating ECG and fundus image data for cardiovascular risk assessment, several limitations must be acknowledged. The lower sensitivity observed in Class 1 (normal ECG + normal fundus) highlights potential difficulties in distinguishing subtle differences between normal and slightly abnormal cases. Our current CNN architecture, while effective, leaves room for improvement, and the model's versatility in detecting a comprehensive range of CVDs needs further validation. The study's limited sample size and geographical scope restrict our ability to assess the model's reliability and generalizability across diverse populations. To address these limitations, we propose expanding the dataset to include a larger, more diverse patient population from multiple geographical regions[40], investigating advanced data augmentation techniques and alternative data fusion methods, and exploring more sophisticated deep learning architectures. Additionally, conducting longitudinal studies and collaborating with clinicians[40] to implement clinical trials will be crucial for validating the model's performance in real-world healthcare settings. By addressing these challenges, we aim to refine our approach and develop a more robust, generalizable, and clinically applicable tool for cardiovascular risk assessment, potentially revolutionizing early detection and management of CVDs in resource-limited healthcare ecosystems.

Future work will focus on several key areas to enhance the model's performance and clinical applicability. First, we plan to explore advanced CNN architectures such as ResNet[41], DenseNet[42], or EfficientNet[43], which have shown promise in medical image analysis. We will also investigate transfer learning techniques, leveraging pre-trained models on large-scale medical imaging datasets to improve our model's feature extraction capabilities. To address the data integration challenge, we will explore more sophisticated fusion techniques such as attention mechanisms or graph neural networks, which could better capture the complex relationships between ECG and fundus image data. Additionally, we aim to develop interpretability tools to provide clinicians with insights into the model's decision-making process, enhancing trust and facilitating clinical adoption. Lastly, we will initiate collaborations with multiple cardiac care centers across different regions to gather a more diverse dataset, including a wider range of cardiovascular conditions and patient demographics. This expanded dataset will enable us to conduct more comprehensive validation studies and assess the model's performance across various subgroups, ultimately leading to a more robust and generalizable tool for cardiovascular risk assessment.

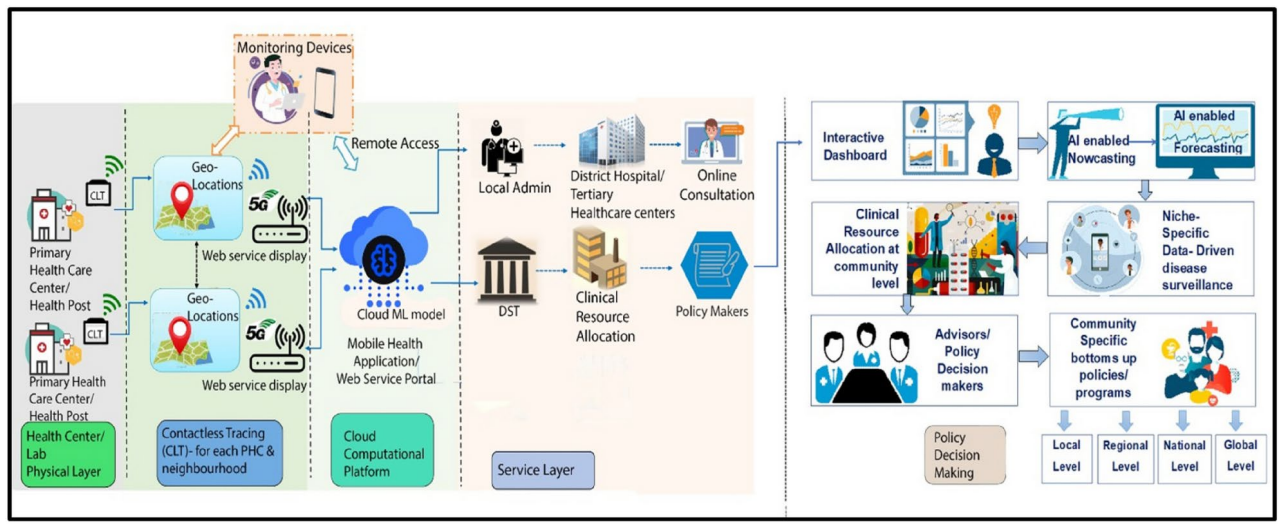

**Fig. 10**. Collaborative AI-driven healthcare system emphasizing the involvement of healthcare providers, patients, and policymakers for effective cardiovascular diagnostics and resource allocation.

For the development of our novel approach, integration of various stakeholders, including but not limited to healthcare providers, patients, and policymakers, is pivotal in the early and precision-oriented detection of CVDs using a combination of multiple modalities, including ECG traces, ECHO recordings, and Retinal Fundus Images (RFIs). In addition to these three modalities, astute clinical inputs from healthcare providers will significantly improve the specificity, sensitivity, and accuracy of the multi-modal, multi-sensor integrated computational platform. This platform is designed for seamless deployment, not only in clinical practice as a point-of-care device but also for facilitating rapid triaging of subjects in community settings (Fig. 10).

Fig. 10 represents a prospective model that illustrates the potential future applications of this platform. While not directly part of the present study, it provides a framework for integrating additional data sources, such as cardiovascular-specific age-, sex-, and race-matched molecular markers, which could enhance the platform's predictive capabilities. This model outlines the ability of the system to identify risk factors and precursors of CVD events at the community level and highlights the rationale for collecting primary data across the 806 districts of the Indian subcontinent. Furthermore, it demonstrates the heuristic capability of the platform to provide automated allocation of clinical resources in collaboration with healthcare administrators and policymakers at the local, regional, national, and global levels. The Fig. 10 emphasizes the long-term vision of the platform in addressing both clinical and community-level challenges. By identifying hotspots of CVDs and facilitating equitable resource distribution, this AI-driven system aims to advance precision diagnostics and triaging, enabling better management and prevention of cardiovascular diseases.

## Contributions of proposed study

- **Superior feature extraction with FFT and EMD**: This study highlights the effectiveness of Fast Fourier Transform (FFT) and Earth Mover's Distance (EMD) in extracting critical features from both ECG signals and fundus images. These methods outperform alternative techniques such as Wavelet Transform (WT) and Histogram of Oriented Gradients (HOG) in cardiovascular diagnostics. The combination of FFT and EMD proved to provide higher accuracy (84%) and sensitivity (88%) in identifying cardiovascular abnormalities, making them the preferred methods for this application.
- **Multi-modal data integration for comprehensive diagnosis**: Our system integrates ECG and fundus images, allowing for a more holistic approach to cardiovascular diagnostics. This multi-modal integration improves diagnostic capabilities, as each data modality provides unique insights into cardiovascular health. Additionally, the future integration of echocardiogram data is expected to further enhance the system's accuracy and robustness, offering structural and functional information about the heart in conjunction with ECG and retinal data.
- **Practical applications in early cardiovascular disease detection**: The developed system has significant potential for early detection and classification of cardiovascular diseases, especially in resource-limited healthcare settings. By integrating data from multiple modalities, the system can provide comprehensive insights into a patient's cardiovascular health, which is critical for early intervention and prevention strategies. This approach can potentially improve clinical outcomes by enabling healthcare providers to make more informed decisions based on multi-source data.
- **Future system refinements with advanced techniques**: In future iterations, the system will incorporate more advanced feature extraction and classification methods. Planned enhancements include the use of visibili-ty graphs for time-series data analysis, eigenvalue-based detection/classification, and Ricci curvature-based classification. These techniques will refine the model's ability to detect subtle patterns, improving both the sensitivity and specificity of the system across a wider range of cardiovascular conditions.

- **Collaboration with healthcare stakeholders for clinical validation**: The study emphasizes the importance of collaboration with healthcare providers, patients, and policymakers to ensure the clinical relevance and applicability of the system. Engaging stakeholders in the development and validation process will help tailor the system to the specific needs of healthcare environments, making it more effective in real-world clinical applications. Future collaborations will involve large-scale validation studies and implementation of the system in clinical trials to assess its performance across diverse populations.
- **Scalability and versatility for broader use**: The system is designed to be scalable and evolvable, allowing for the integration of new data types and advanced techniques as they emerge. This ensures that the system can adapt to new developments in cardiovascular diagnostics and remain relevant across different healthcare settings. The integration of cloud-based platforms for data processing further enhances the scalability of the system, making it deployable across local, regional, national, and global healthcare ecosystems.

## Conclusion

In conclusion, our study presents a novel approach to cardiovascular risk assessment by integrating ECG, and fundus image data using advanced machine learning techniques, demonstrating promising results with an overall accuracy of 84%. This multi-modal approach, combining Earth Mover's Distance for data fusion and Convolutional Neural Networks for classification, offers a unique perspective on leveraging diverse clinical data for improved diagnostics. However, we acknowledge limitations in our current model, including sensitivity challenges for normal cases, data integration complexities, and the need for broader validation across diverse populations. These limitations guide our future research directions, which include expanding our dataset, refining analytical methods, and exploring more sophisticated neural network architectures. The potential impact of this work is particularly significant for resource-limited healthcare ecosystems, where it could serve as a valuable tool for early screening and risk stratification. As we continue to refine and validate this model through expanded studies and clinical trials, we envision a future where integrated, AI-assisted cardiovascular diagnostics become an integral part of routine healthcare. While our current results are encouraging, they represent a stepping stone towards a more comprehensive, accurate, and accessible approach to cardiovascular health assessment. Through collaborative efforts across medical institutions, continued technological innovation, and rigorous clinical validation, we aspire to contribute significantly to the global fight against cardiovascular diseases, potentially transforming the landscape of cardiovascular disease prevention and management through earlier detection and more personalized interventions. The four-class framework developed in this study represents a significant advancement in integrating ECG and fundus imaging for cardiovascular diagnostics. By going beyond binary classification, the study uncovers nuanced patterns that improve early detection and preventive strategies. This approach enables more targeted patient triaging and highlights the importance of multimodal diagnostic platforms in resource-limited healthcare settings.

## Data availability

The datasets used and analysed during the current study are available from the corresponding author on reasonable request.

## References
1. Cardiovascular diseases (CVDs) — WHO.int. https://www.who.int/news-room/fact-sheets/detail/cardiovascular-diseases-(cvds)
2. Wang, H. et al. Global, regional, and national life expectancy, all-cause mortality, and cause-specific mortality for 249 causes of death, 1980–2015: A systematic analysis for the global burden of disease study 2015. *The lancet* **388**(10053), 1459–1544 (2016).
3. Goff Jr, D.C., Lloyd-Jones, D.M., Bennett, G., Coady, S., D'agostino, R.B., Gibbons, R., Greenland, P., Lackland, D.T., Levy, D., O'donnell, C.J., *et al.*: 2013 ACC/AHA guideline on the assessment of cardiovascular risk: A report of the american college of cardiology/american heart association task force on practice guidelines. Circulation 129(25_suppl_2), 49–73 (2014)
4. Iqbal, S., Khan, T. M., Naveed, K., Naqvi, S. S. & Nawaz, S. J. Recent trends and advances in fundus image analysis: A review. *Comput. Biol. Med.* **151**, 106277 (2022).
5. Goutam, B., Hashmi, M. F., Geem, Z. W. & Bokde, N. D. A comprehensive review of deep learning strategies in retinal disease diagnosis using fundus images. *IEEE Access* **10**, 57796–57823 (2022).
6. Farrah, T. E., Dhillon, B., Keane, P. A., Webb, D. J. & Dhaun, N. The eye, the kidney, and cardiovascular disease: Old concepts, better tools, and new horizons. *Kidney Int.* **98**(2), 323–342 (2020).
7. Wong, T. Y. et al. Retinal microvascular abnormalities and their relationship with hypertension, cardiovascular disease, and mortality. *Surv. Ophthalmol.* **46**(1), 59–80 (2001).
8. Kooi, T. et al. Large scale deep learning for computer aided detection of mammographic lesions. *Med. Image Anal.* **35**, 303–312 (2017).
9. Ghafoorian, M., Karssemeijer, N., Heskes, T., Van Uder, I., Leeuw, F.-E., Marchiori, E., Ginneken, B., Platel, B.: Non-uniform patch sampling with deep convolutional neural networks for white matter hyperintensity segmentation. in *2016 IEEE 13th International Symposium on Biomedical Imaging (ISBI)*, IEEE, pp. 1414–1417 (2016)
10. Esteva, A. et al. Dermatologist-level classification of skin cancer with deep neural networks. *Nature* **542**(7639), 115–118 (2017).
11. Ozturk, T. et al. Automated detection of covid-19 cases using deep neural networks with x-ray images. *Comput. Biol. Med.* **121**, 103792 (2020).
12. Rubner, Y., Tomasi, C. & Guibas, L. J. The earth mover's distance as a metric for image retrieval. *Int. J. Comput. Vision* **40**, 99–121 (2000).
13. Flammer, J. et al. The eye and the heart. *Eur. Heart J.* **34**(17), 1270–1278 (2013).

14. Oloumi, F., Rangayyan, R.M., Ells, A.L.: Digital image processing for ophthalmology: Detection and modeling of retinal vascular architecture. Morgan & Claypool Publishers (2014)
15. Huang, L. et al. Exploring associations between cardiac structure and retinal vascular geometry. *J. Am. Heart Assoc.* **9**(7), 014654 (2020).
16. McGeechan, K. et al. Meta-analysis: Retinal vessel caliber and risk for coronary heart disease. *Ann. Intern. Med.* **151**(6), 404–413 (2009).
17. Seidelmann, S. B. et al. Retinal vessel calibers in predicting long-term cardiovascular outcomes: The atherosclerosis risk in communities study. *Circulation* **134**(18), 1328–1338 (2016).
18. Arnould, L. et al. Association between the retinal vascular network with singapore "i" vessel assessment (siva) software, cardiovascular history and risk factors in the elderly: The montrachet study, population-based study. *PLoS ONE* **13**(4), 0194694 (2018).
19. Wong, T. Y. et al. Atherosclerosis risk in communities study. *The Lancet* **358**, 1134–40 (2001).
20. Wong, T. Y. et al. Retinal arteriolar narrowing and risk of coronary heart disease in men and women: The atherosclerosis risk in communities study. *JAMA* **287**(9), 1153–1159 (2002).
21. Wong, T. Y. et al. Retinal microvascular abnormalities and 10-year cardiovascular mortality: A population-based case-control study. *Ophthalmology* **110**(5), 933–940 (2003).
22. Armstrong, G. W. & Lorch, A. C. A (eye): A review of current applications of artificial intelligence and machine learning in ophthalmology. *Int. Ophthalmol. Clin.* **60**(1), 57–71 (2020).
23. Desai, U. et al. Decision support system for arrhythmia beats using ECG signals with DCT, DWT and Empirical Mode Decomposition methods: A comparative study. *J. Mech. Med. Biol.* **16**(1), 1640012 (2016).
24. Desai, U., Nayak, C. G. & Seshikala, G. Application of ensemble classifiers in accurate diagnosis of myocardial ischemia conditions. *Progr. Artif. Intell.* **6**, 245–253 (2017).
25. Poplin, R. et al. Prediction of cardiovascular risk factors from retinal fundus photographs via deep learning. *Nat. Biomed. Eng.* **2**(3), 158–164 (2018).
26. Krittanawong, C. et al. Deep learning for cardiovascular medicine: A practical primer. *Eur. Heart J.* **40**(25), 2058–2073 (2019).
27. Cheung, C. Y. et al. A deep-learning system for the assessment of cardiovascular disease risk via the measurement of retinal-vessel calibre. *Nat. Biomed. Eng.* **5**(6), 498–508 (2021).
28. Niccoli, T. & Partridge, L. Ageing as a risk factor for disease. *Curr. Biol.* **22**(17), 741–752 (2012).
29. Vaghefi, E., Squirrell, D., Yang, S., An, S., Marshall, J.: Use of artificial intelligence on retinal images to accurately predict the risk of cardiovascular event (cvd-ai). medRxiv, 2022–10 (2022)
30. Tseng, R. M. W. W. et al. Validation of a deep-learning-based retinal biomarker (reti-cvd) in the prediction of cardiovascular disease: Data from uk biobank. *BMC Med.* **21**(1), 1–10 (2023).
31. Diaz-Pinto, A. et al. Predicting myocardial infarction through retinal scans and minimal personal information. *Nat. Mach. Intell.* **4**(1), 55–61 (2022).
32. Nandi, D. et al. A multi-modal computational approach to accurate and rapid prediction of cardiovascular health using fundus images. *Ind. Heart J.* **75**, 9. https://doi.org/10.1016/j.ihj.2023.11.027 (2023).
33. Kingsbury, N. & Magarey, J. *Wavelet transforms in image processing* (Springer, 1998).
34. Dalal, N., Triggs, B.: Histograms of oriented gradients for human detection. In: *2005 IEEE Computer Society Conference on Computer Vision and Pattern Recognition (CVPR'05)*. IEEE, vol. 1, pp. 886–893 (2005)
35. Muthukumar, K., Bouazizi, M. & Ohtsuki, T. A novel hybrid deep learning model for activity detection using wide-angle low-resolution infrared array sensor. *IEEE Access* **9**, 82563–82576 (2021).
36. Krishnan, A. M., Bouazizi, M. & Ohtsuki, T. An infrared array sensor-based approach for activity detection, combining low-cost technology with advanced deep learning techniques. *Sensors* **22**(10), 3898 (2022).
37. Hubbard, L.D., Brothers, R.J., King, W.N., Clegg, L.X., Klein, R., Cooper, L.S., Sharrett, A.R., Davis, M.D., Cai, J., Communities Study Group, A.R., et al.: Methods for evaluation of retinal microvascular abnormalities associated with hypertension/sclerosis in the atherosclerosis risk in communities study. Ophthalmology, 106(12):2269–2280 (1999)
38. Huang, F., Lian, J., Ng, K.-S., Shih, K. & Vardhanabhuti, V. Predicting ct-based coronary artery disease using vascular biomarkers derived from fundus photographs with a graph convolutional neural network. *Diagnostics* **12**(6), 1390 (2022).
39. Srilakshmi, V., Anuradha, K. & Bindu, C. S. Intelligent decision support system for cardiovascular risk prediction using hybrid loss deep joint segmentation and optimized deep learning. *Adv. Eng. Softw.* **173**, 103198 (2022).
40. Ganeshan, M. A. et al. Multi-modal detection of cardiovascular disease at the community level. *Ind. Heart J.* **75**, 76–77. https://doi.org/10.1016/j.ihj.2023.11.162 (2023).
41. Sakli, N. et al. Resnet-50 for 12-lead electrocardiogram automated diagnosis. *Comput. Intell. Neurosci.* **2022**(1), 7617551 (2022).
42. Krishnan, V. G. et al. Hybrid optimization based feature selection with densenet model for heart disease prediction. *Int. J. Electr. Electron Res.* **11**, 253–61 (2023).
43. Han, C. & Yoon, D. An explainable artificial intelligence-enabled ECG framework for the prediction of subclinical coronary atherosclerosis. *AMIA Summ. Translat. Sci. Proceed.* **2024**, 535 (2024).

## Author contributions

Muthukumar KA- Conceptualization, Methodology, building models, Implementation, Manuscript writing, and reviewing. Dhruva Nandi- Conceptualization, data collection and writing. Priya Ranjan- Conceptualization reviewing. Krithika Ramachandran- writing and reviewing. Shiny PJ- Data collection and reviewing. Anirban Ghosh-Data collection and reviewing. Ashwini M- Data collection (fundus) and supervision. Aiswaryah Radhakrishnan- Data collection (fundus) and supervision. Dhandapani VE- Data collection (ECGs) and supervision. Rajiv Janardhanan- Conceptualization, reviewing and supervision.

## Funding

Open access funding provided by SRM Institute of Science and Technology for SRMIST – Medical & Health Sciences.

No funding was obtained for this study.

## Declarations

### Competing interests
The authors declare no competing interests.

### Ethical approval
A proper written informed consent has been obtained before the commencement of the study.

### Consent for publication
The publisher has the author's permission to publish research findings.

### Additional information
**Correspondence** and requests for materials should be addressed to K.A.M. or R.J.

**Reprints and permissions information** is available at www.nature.com/reprints.

**Publisher's note**  Springer Nature remains neutral with regard to jurisdictional claims in published maps and institutional affiliations.